\documentclass[lettersize,conference]{IEEEtran}
\IEEEoverridecommandlockouts
\usepackage{geometry}
\usepackage{textcomp}
\usepackage{xcolor}

\usepackage{cite}
\usepackage{amsmath,amssymb,amsfonts}
\usepackage{graphicx,subfigure,epstopdf}
\usepackage{textcomp}
\usepackage[colorlinks,linkcolor=black]{hyperref}
\usepackage{algorithm}
\usepackage{algorithmicx}
\usepackage{algpseudocode}  
\usepackage{amsmath}  
\usepackage{multirow}
\usepackage[T1]{fontenc} 

\usepackage{amsthm} % 如果要使用proof语句就必须要引入这个包

\def\BibTeX{{\rm B\kern-.05em{\sc i\kern-.025em b}\kern-.08em
    T\kern-.1667em\lower.7ex\hbox{E}\kern-.125emX}}

\begin{document}
\title{Minimizing Movement Delay for Movable Antennas via Trajectory Optimization}

\author{\IEEEauthorblockN{Qingliang Li\IEEEauthorrefmark{1}, Weidong Mei\IEEEauthorrefmark{1}, Boyu Ning\IEEEauthorrefmark{1}, and Rui Zhang\IEEEauthorrefmark{2}\IEEEauthorrefmark{3} }
	\IEEEauthorblockA{\IEEEauthorrefmark{1} \textit{National Key Laboratory of Wireless Communications}\\
	\textit{University of Electronic Science and Technology of China (UESTC),
		Chengdu, China.} \\
	\IEEEauthorrefmark{2} \textit{School of Science and Engineering, Shenzhen Research Institute of Big Data} \\
	\textit{Chinese University of Hong Kong, Shenzhen, China. } \\
	\IEEEauthorrefmark{3} \textit{Department of Electrical and Computer Engineering, National University of Singapore. }\\
	\textit{Emails: liqingliang@std.uestc.edu.cn; wmei@uestc.edu.cn; boydning@outlook.com; elezhang@nus.edu.sg}
		\vspace{-1cm}
	%	e-mail: liqingliang@std.uestc.edu.cn; changb3212@163.com; wmei@uestc.edu.cn; chenzhi@uestc.edu.cn
}

%\thanks{The authors are with the National Key Laboratory of Wireless Communications, University of Electronic Science and Technology of China (UESTC), Chengdu, 611731, China. (e-mail: liqingliang@std.uestc.edu.cn; changb3212@163.com; wmei@uestc.edu.cn; chenzhi@uestc.edu.cn)}

}

\maketitle
\begin{abstract}
	Movable antennas (MAs) have received increasing attention in wireless communications due to their capability of antenna position adjustment to reconfigure wireless channels. However, moving MAs results in non-negligible delay, which may decrease the effective data transmission time. To reduce the movement delay, we study in this paper a new MA trajectory optimization problem. In particular, given the desired destination positions of multiple MAs, we aim to jointly optimize their associations with the initial MA positions and the trajectories for moving them from their respective initial to destination positions within a given two-dimensional (2D) region, such that the delay of antenna movement is minimized, subject to the inter-MA minimum distance constraints in the movement. However, this problem is a continuous-time mixed-integer linear programming (MILP) problem that is challenging to solve. To tackle this challenge, we propose a two-stage optimization framework that sequentially optimizes the MAs' position associations and trajectories, respectively. First, we relax the inter-MA distance constraints and optimally solve the resulted delay minimization problem. Next, we check if the obtained MA association and trajectory solutions satisfy the inter-MA distance constraints. If not satisfied, we then employ a successive convex approximation (SCA) algorithm to adjust the MAs' trajectories until they satisfy the given constraints. 
Simulation results are provided to show the effectiveness of our proposed trajectory optimization method in reducing the movement delay as well as draw useful insights.

\end{abstract}
%

%\begin{IEEEkeywords}
%Movable antenna, trajectory optimization, successive convex approximation.
%\end{IEEEkeywords}

\section{Introduction}
Recently, movable antennas (MAs) have garnered significant attention from both academia and industry in the realm of wireless communications. By dynamically adjusting their positions and/or rotations within a confined spatial region, MAs can adaptively refine the wireless channel condition in favor of signal transmission \cite{MCom2024_Zhu,shao_6DMA}. Compared to conventional fixed-position antennas, MAs offer additional degrees of freedom for enhancing wireless communication performance, which opens up new research opportunities for future wireless networks.

Inspired by the benefits of MAs, a variety of recent works have investigated their performance optimization problems under various setups, e.g., single- and multi-user multiple-input single-output (MISO) \cite{MA_Mei,TWC2024_Zhu}, multi-input multi-output (MIMO) \cite{TWC2024_Ma}, physical-layer security \cite{MA_secure}, cognitive radio \cite{Wei2024}, interference channel\cite{wang2024movable}, etc. However, MAs are usually implemented with mechanical drivers such as motor as compared to other methods to enable antenna movement such as liquid and deployable antennas  \cite{ning2024movable}. For instance, motor-enabled MAs have been utilized in a multi-static radar system in \cite{MA1} to move both the transmit and receive antennas along parallel paths. 
Furthermore, the authors in \cite{FA} employed liquid metal to enable antenna movement within a container by applying voltage to electrodes, which drives the antenna through the Marangoni force. The authors in \cite{DA} leverage the internal mechanical structure of deployable antennas (also known as origami antenna) to alter the geometry of multiple antenna arrays.
However, mechanically driven MAs typically result in non-negligible delay for moving MAs. Furthermore, the movement of one antenna may be constrained by the positions of others to avoid collisions or mutual coupling, thus further increasing their movement delay. As a result, the effective data transmission time may be comprised given a finite channel coherence time.

\begin{figure}[!t]
	\centering
	\includegraphics[scale=0.75]{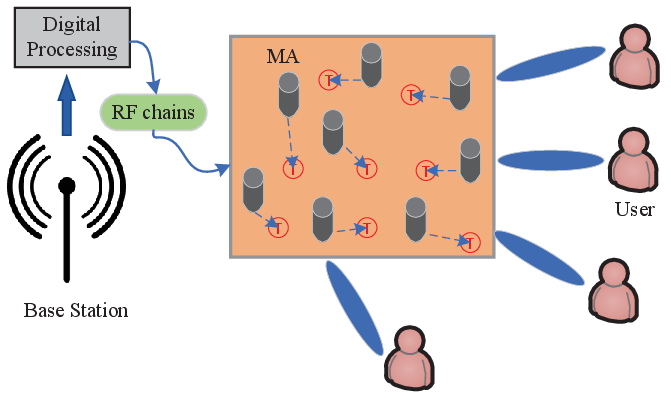}
	\setlength{\abovecaptionskip}{-0.2cm}
	\setlength{\belowcaptionskip}{-0.2cm} %调整图片标题与下文距离
	\caption{MA-assisted wireless communications with a trajectory design.}
	\vspace{-0.1cm}
	\label{fig:MA}
\end{figure}

In this paper, we aim to resolve the above issue from the perspective of MA trajectory optimization. To the best of our knowledge, this paper is the first to study the MA trajectory optimization problem in the literature. In particular, as shown in Fig. \ref{fig:MA}, for a typical MA-enhanced wireless communication system with desired destination positions of multiple MAs, we aim to jointly optimize their associations with the initial MA positions and the trajectories for moving them from their respective initial to destination positions within a given two-dimensional (2D) region, such that the delay of antenna movement is minimized, subject to the inter-MA minimum distance constraints in the movement. However, this problem is a continuous-time mixed-integer linear programming (MILP) problem that is difficult to solve. To tackle this difficulty, we propose a two-stage optimization framework by sequentially optimizing the MAs' position associations and trajectories, respectively. Specifically, we first relax the inter-MA distance constraints and derive the corresponding position association and trajectory solutions by optimally solving a simplified MILP. Next, we check if these solutions can satisfy the inter-MA distance constraints. If not satisfied, we then propose a successive convex approximation (SCA) algorithm to adjust the MAs' trajectories until they satisfy the given constraints. 
Simulation results show the effectiveness of our proposed trajectory optimization method in reducing the movement delay and drive essential insights into the optimal trajectory design.

{\it Notations}: Bold symbols in lowercase and uppercase denote vectors and matrices, respectively. $\mathbb{R}^{n \times m}$ denotes the set of all $n \times m$ real matrices. The uppercase letters in calligraphy fonts are utilized to denote sets, such as $\mathcal{M}$.
$\Vert \cdot \Vert$ represents the Euclidean distance. $\boldsymbol{p}^\top$ denote the transpose of a vector $\boldsymbol{p}$. 
%$(a,A)$ indicates that antenna $a$ is assigned to target position $A$.

\section{System Model and Problem Formulation}

\begingroup
\allowdisplaybreaks

\begin{figure}[!t]
	\centering
	\includegraphics[scale=0.9]{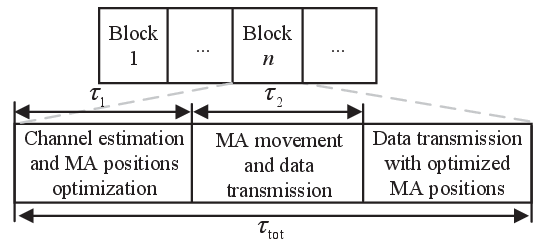}
	\setlength{\abovecaptionskip}{-0.2cm}
	\setlength{\belowcaptionskip}{-0.2cm} %调整图片标题与下文距离
	\caption{Transmission protocol of MA-assisted wireless communication.}
	\vspace{-0.1cm}
	\label{fig:frame}
\end{figure}
As shown in Fig. \ref{fig:MA}, we consider a general MA-assisted multi-user communications system, where a base station (BS) equipped with $M$ MAs serves multiple users at the same time. We assume that all MAs can be flexibly moved within a 2D region, denoted as ${\cal C}_m$. We consider block-fading channels between the BS and all users and focus on one block for the proposed MA movement trajectory design. As shown in Fig. \ref{fig:frame}, at the beginning of each transmission block, the BS first estimates the BS-user channels for any given position in ${\cal C}_m$,\footnote{This can be achieved by applying the compressed-sensing-based channel estimation schemes catered to MAs (see e.g., \cite{Channel} and \cite{Channel2})} based on which it optimizes the beamforming and positions for all MAs. Then, the MAs are moved to the optimized positions for improving the data transmission rate.  Let $\tau_{\text{tot}}$, $\tau_1$, and $\tau_2$ denote the total duration of each transmission block, the duration for channel estimation, and that for MA movement, respectively.
As such, the effective achievable rate of any user (e.g., user $k$) in bits/second/Hertz (bps/Hz) is given by
\begin{align}\label{rate}
	R_k \geq \left( 1-\frac{\tau_1+\tau_2}{\tau_{\text{tot}}} \right) {\hat R}_k,
\end{align}
where ${\hat R}_k$ is the achievable rate with the optimized MA positions. Note that the inequality in \eqref{rate} holds because there is still data transmission (with lower rates than ${\hat R}_k$ in general) during the MA movement, while it becomes equality if the antenna movement duration $\tau_2 \rightarrow 0$. As the length of $\tau_1$ depends on the channel estimation scheme adopted for MAs (which is not the focus of our work), the achievable rate given in \eqref{rate} also critically depends on $\tau_2$ for MA movement. In particular, a longer/shorter duration of $\tau_2$ results in a smaller/larger achievable rate in \eqref{rate}. Inspired by this, we study in this paper the trajectory optimization problem for MAs to minimize their movement delay, $\tau_2$.

To this end, we first discretize the duration $\tau_2$ into $N$ time slots, each with a length of $\tau$. We assume that $\tau$  is sufficiently small, such that the position of each MA can be approximately viewed as being constant within each time slot. Let $\boldsymbol{p}_m[n] \in {\mathbb R}^{2 \times 1}$ denote the position of MA $m$ in ${\cal C}_m$ in time slot $n$, with $m\in \mathcal{M} \triangleq \{1,2,\cdots,M\}$, and $n\in \mathcal{N} \triangleq \{0,1,2,\cdots,N\}$. 
To ensure that the MAs do not collide with each other or cause mutual coupling in the movement, we consider the following constraints on the minimum inter-MA distance allowed, i.e.,
\begin{equation}\label{con_d}
	d_{\min} \leq \Vert \boldsymbol{p}_m [n] - \boldsymbol{p}_j [n] \Vert,\ m \neq j,\ m\in \mathcal{M},\ j\in \mathcal{M},\ n\in \mathcal{N},
\end{equation}
where $d_{\min}$ is the given minimal distance between any two MAs in the movement.
Moreover, due to the finite power to drive MAs (e.g., via a motor), we consider a maximum speed for each MA in the movement, i.e.,
\begin{equation}\label{con_v}
	\Vert \boldsymbol{p}_m [n+1] - \boldsymbol{p}_m [n] \Vert \leq V_{\max}\tau,\  m\in \mathcal{M},\ n\in \mathcal{N},
\end{equation}
where $V_{\max}$ is the maximum speed of each MA.

Let the initial position of MA $m$, $m \in {\cal M}$, be denoted as ${\boldsymbol{p}}_{m,0} \in {\mathbb R}^{2 \times 1}$ and ${\mathcal{P}}_0 \triangleq \{{\boldsymbol{p}}_{1,0}, {\boldsymbol{p}}_{2,0}, \cdots, {\boldsymbol{p}}_{M,0} \}$ denote the set of the initial positions of all $M$ MAs. 
Evidently, we have $\boldsymbol{p}_m[0]=\boldsymbol{p}_{m,0}$, $\forall m \in {\cal M}$. 
Moreover, let the $j$-th desired destination position $j$, $j \in \mathcal{M}$, be denoted as $\boldsymbol{q}_j \in {\mathbb R}^{2 \times 1}$ and $\mathcal{Q} \triangleq \{\boldsymbol{q}_1, \boldsymbol{q}_2, \cdots, \boldsymbol{q}_M\}$ denote the set of the desired destination positions of all $M$ MAs. 
It is noted that a new initial-destination position antenna association problem arises for matching each $\boldsymbol{p}_{m,0}$ in ${\mathcal{P}}_0$ with one of the destination positions in $\cal Q$. In particular, an improper association may result in large movement delay. For example, as shown in Fig. \ref{fig:MA}, the leftmost MA at its initial position would experience a much longer delay if it is moved to the rightmost destination position. To describe the MA position associations, we introduce a set of binary variables $\sigma_{m,j}, m, j \in {\cal M}$. If $ \sigma_{m,j} = 1 $, it implies that the destination position $\boldsymbol{q}_j$ is associated with MA $ m $ (or $\boldsymbol{p}_{m,0}$); otherwise, it is equal to zero. Since each initial/destination position can be associated with exactly one destination/initial position, it must hold that $\sum_{m=1}^{M} \sigma_{m,j} = 1,\ j \in \mathcal{M}$ and $\sum_{j=1}^{M} \sigma_{m,j} = 1,\ m \in \mathcal{M}$.

Based on the above, we aim to jointly optimize the trajectories of the $M$ MAs (i.e., $\boldsymbol{P} = \{\boldsymbol{p}_m [n],\ m \in \mathcal{M},\ n\in \mathcal{N}\}$), the MA position associations (i.e., $\boldsymbol{A} = \{\sigma_{m,j},\ m \in \mathcal{M},\ j \in \mathcal{M} \}$), and the duration of each time slot (i.e., $\tau$) to minimize the overall movement delay, i.e., $\tau_2=N\tau$, (assuming that $N$ is fixed). The corresponding optimization problem is formulated as

\begin{align}\label{eq:P_1}
		{\text{(P1)} }&:  \quad \min_{\boldsymbol{P}, \boldsymbol{A}, \tau} \ \quad  \tau_2=N\tau \\
		s.t.&\quad  \eqref{con_d},\ \eqref{con_v}, \nonumber \\ 
		&\quad \boldsymbol{p}_m [n] \in {\cal C}_m,\ m \in \mathcal{M},\ n \in \mathcal{N}, \label{con_boundary} \\
		&\quad \boldsymbol{p}_m [0]=\boldsymbol{p}_{m,0},\ m \in \mathcal{M}, \label{con_p0} \\
		&\quad \sigma_{m,j} \Vert \boldsymbol{p}_m [N] - \boldsymbol{q}_{j}\Vert = 0,\ m \in \mathcal{M},\ j \in \mathcal{M}, \label{con_pN}\\
		&\quad \sigma_{m,j} \in \{0,1\},\ m \in \mathcal{M},\ j \in \mathcal{M}, \label{con_a} \\
		&\quad \sum_{m=1}^{M} \sigma_{m,j} = 1,\ j \in \mathcal{M}, \label{con_am}\\ 
		&\quad \sum_{j=1}^{M} \sigma_{m,j} = 1,\ m \in \mathcal{M}, \label{con_aj}
\end{align}
where the constraints in \eqref{con_pN} ensure that MA $m$ can be moved to its associated destination position by time slot $N$. However, it is noted that (P1) is an NP-hard MILP problem due to the integer association variables $\{\sigma_{m,j}\}$ and the non-convex constraints in \eqref{con_d}. Next, we will propose an efficient two-stage algorithm to solve it.

\section{Proposed Solution to (P1)}
In this section, we present the details of our proposed algorithm for solving (P1).

\subsection{Optimal Solution to (P1) Without Inter-MA Distance Constraints}
First, we ignore the inter-MA distance constraints in \eqref{con_d}. In this case, to achieve the minimum movement delay, it is evident that each MA should be moved along a straight path from its initial  position to its associated destination position at the maximum speed, $V_{\max}$. It can be shown that the resulting movement delay is given by
\begin{align}\label{T_lb}
	\hat\tau_2 = \min_{\sigma_{m,j}}  \frac{\max_{m, j\in \mathcal{M}} \sigma_{m,j} \Vert \boldsymbol{p}_{m,0} - \boldsymbol{q}_{j}\Vert}{V_{\max}},
\end{align}
in terms of the association variables in $\boldsymbol{A}$. Note that \eqref{T_lb} serves as a lower bound on the optimal value of (P1) and can be obtained by solving a simplified MILP shown in \eqref{T_lb} compared to (P1). Let 
\begin{align}
	\mathcal{D} = \left\{d_{m,j} = \Vert \boldsymbol{p}_{m,0} - \boldsymbol{q}_{j}\Vert,\ m \in \mathcal{M},\ j\in \mathcal{M} \right\},
\end{align}
denote the set of distances between the initial  and destination MA positions. Then, the optimal association solution that yields $\hat\tau_2$ can be obtained by solving the following MILP,
\begin{align}\label{eq:P_2}
	{\text{(P2)} }&:  \quad \min_{\boldsymbol{A}} \ \max_{m,j\in \mathcal{M}} \quad \sigma_{m,j} d_{m,j} \\
	s.t.
	&\quad \eqref{con_a},\ \eqref{con_am},\ \eqref{con_aj}, \nonumber 
\end{align}
where we have ignored the constant scalar $V_{\max}$ in (P2). For (P2), we can further recast it as an epigraph form by introducing a slack variable $\xi$, i.e.,
\begin{align}\label{eq:P_21}
	{\text{(P2.1)} }&:  \quad \min_{\boldsymbol{A}, \xi} \ \quad \xi \\
	s.t. 	&\quad \eqref{con_a},\ \eqref{con_am},\ \eqref{con_aj}, \nonumber \\
	&\quad \sigma_{m,j} d_{m,j} \leq \xi,\ m \in \mathcal{M},\ j \in \mathcal{M}. \label{con_xi}
\end{align}
Problem (P2.1) is a standard MILP, which can be optimally solved by employing the branch-and-bound algorithm via the off-the-shelf MATLAB solver {\it intlinprog} \cite{MATLABintlinprog}.

Let $\boldsymbol{A}^*=\{\sigma^*_{m,j},\ m \in \mathcal{M},\ j \in \mathcal{M}\}$ and $\xi^*$ denote the optimal solutions to (P2.1). As such, if the constraints in \eqref{con_d} are ignored, MA $m$, $m \in {\cal M}$, should be moved from $\boldsymbol{p}_{m,0}$ to $\boldsymbol{p}_{j(m)}$ with the maximum speed of $V_{\max}$, with $j(m)=\{j | j \in {\cal M}, \sigma^*_{m,j}=1\}$. 
The length of each time slot is given by ${\hat\tau}=\xi^*/(V_{\max}N)$. The position of MA $m$ in time slot $n$ is thus given by
\begin{align}\label{eq:Vm}
	\hat{\boldsymbol{p}}_m[n] =
	\left\{ 
	\begin{array}{ll}
		\boldsymbol{p}_m [0] + \frac{{\boldsymbol{q}}_{j(m)}-{\boldsymbol{p}}_{m,0}}{\Vert {\boldsymbol{q}}_{j(m)}-{\boldsymbol{p}}_{m,0} \Vert} V_{\max} n{\hat\tau}, &\text{ if}\ n \le \hat{N},\\
		{\boldsymbol{q}}_{j(m)}, &\text{ otherwise}.
	\end{array} \right.
\end{align}
where $\hat{N} = \lfloor \frac{N\lVert \boldsymbol{q}_{j(m)}-\boldsymbol{p}_{m,0} \rVert}{\xi^*} \rfloor$.

\subsection{Feasibility Check and Modified Trajectory}
However, the association solution $\sigma^*_{m,j}$ and the obtained trajectory solution in \eqref{eq:Vm} may not be feasible to (P1) when the inter-MA distance constraints in \eqref{con_d} are applied. Next, we check if these obtained solutions are feasible for (P1). To this end, for each time slot $n$, we calculate $\lVert \boldsymbol{p}_m[n] - \boldsymbol{p}_j[n] \rVert$ for any $m$ and $j$, $m \ne j$. If the following conditions hold, i.e.,
\begin{equation}\label{eq:check}
	\lVert \boldsymbol{p}_m[n] - \boldsymbol{p}_j[n] \rVert \ge d_{\min}, \forall m \ne j, m,j \in {\cal M}, n \in {\cal N},
\end{equation}
then it can be declared that \eqref{eq:Vm} is a feasible and also a globally optimal solution to (P1). Otherwise, we need to modify the solution in \eqref{eq:Vm} to obtain a feasible solution. In this case, to approach the lower bound in \eqref{T_lb}, we propose to fix the initial-destination position associations as $\sigma_{m,j}^*$ and only modify the trajectory solution in \eqref{eq:Vm}. Accordingly, the associated problem  can be formulated by replacing \eqref{con_pN} in (P1) with $\boldsymbol{p}_m [N] = {\boldsymbol{q}}_{j(m)},\ m \in \mathcal{M}$,  i.e.,
\begin{align}
	{\text{(P3)} }&:  \quad \min_{\boldsymbol{P}, \tau} \ \quad \tau_2=N\tau \nonumber \\
	s.t. &\quad  \eqref{con_d},\ \eqref{con_v}, \eqref{con_boundary}, \nonumber \\ 
	&\quad \boldsymbol{p}_m [0]={\boldsymbol{p}}_{m,0},\ m \in \mathcal{M}, \label{con_p1} \\
	&\quad \boldsymbol{p}_m [N] = {\boldsymbol{q}}_{j(m)},\ m \in \mathcal{M}. \label{con_pN1}
\end{align}

Problem (P3) is still a non-convex optimization problem, due to the non-convex constraints in \eqref{con_d}. To deal with this non-convexity, we employ the SCA algorithm. Let $\boldsymbol{P}^{(l)} = \{\boldsymbol{p}_m^{(l)} [n],\ m \in \mathcal{M},\ n\in \mathcal{N}\}$ denote the local trajectories of the MAs in the $l$-th SCA iteration. We implement the first-order Taylor expansion on the square of the right-hand side of \eqref{con_d} and obtain its lower bound as follows: 
\begin{align}
	\Vert \boldsymbol{p}_m [n] - \boldsymbol{p}_j [n] \Vert^2 \geq  Q_{lb}^{(m,j,l)} [n],
\end{align}
where $Q_{lb}^{(m,j,l)} [n] = -\Vert \boldsymbol{p}_m^{(l)} [n] - \boldsymbol{p}_j^{(l)} [n] \Vert^2 + 2 \left(\boldsymbol{p}_m^{(l)} [n] - \boldsymbol{p}_j^{(l)} [n] \right)^\top \left(\boldsymbol{p}_m [n] - \boldsymbol{p}_j [n] \right)$.
By this means, we can replace \eqref{con_d} with the following inequality
\begin{align}\label{con_cvx_d}
	d_{\min}^2 \leq  Q_{lb}^{(m,j,l)} [n],
\end{align}
in the $l$-th SCA iteration and solve the following problem, 
\begin{align}
	{\text{(P3.$l$)} }&:  \min_{\boldsymbol{P}, \tau} \quad \tau_2 =N\tau \nonumber \\
	s.t.&\quad  \eqref{con_v},\ \eqref{con_boundary},\ \eqref{con_p1},\ \eqref{con_pN1},\ \eqref{con_cvx_d}. \nonumber 
\end{align}
Problem (P3.$l$) is convex and can be solved by the interior-point algorithm via the off-the-shelf Matlab toolbox \textsc{CVX} \cite{cvx}. Next, we update $\boldsymbol{P}^{(l+1)}$ as the optimal solution to (P3.$l$) and proceed to solve (P3.$l+1$). Let $\tau_2^{(l)}$ denote the optimal value of (P3) with ${\boldsymbol{P}}=\boldsymbol{P}^{(l)}$. It can be shown that $\tau_2^{(l)}$ should not increase with $l$ and thus, convergence is ensured to be reached \cite{SCA}. The initial MA trajectory, i.e., $\boldsymbol{P}^{(1)}$, can be set as \eqref{eq:Vm}, with $\tau_2^{(1)}=\infty$. We summarize the main procedures of our proposed two-stage optimization algorithm for solving (P1) in Algorithm \ref{alg:SCA}. It is worth noting that the movement delay by the modified MA trajectory may be still the same as that by \eqref{eq:Vm} without any modification. This is because the movement delay $\tau_2$ is constrained by the MA with the longest movement time among all MAs, while its trajectory may not need to be modified. In this case, the modified MA trajectories are still optimal solutions to (P1).

\begin{algorithm}
	\caption {Proposed Two-Stage Algorithm for Solving (P1)}
	\label{alg:SCA} 
	\begin{algorithmic}[1]
		\State Solve problem (P2.1) to obtain $\boldsymbol{A}^*$ and $\xi^*$;
		\State Construct the initial trajectory $\boldsymbol{P}^{(1)}$ based on \eqref{eq:Vm};
		\State Check the feasibility of $\boldsymbol{P}^{(1)}$ based on \eqref{eq:check};
		\State Let $l = 1$, $\tau_2^{(l)}=\infty$, and $\epsilon=\infty$;
		\If{$\boldsymbol{P}^{(1)}$ is infeasible}
		\While{$\epsilon > \epsilon^{*}$}
		\State Solve problem (P3.$l$) to obtain the optimal trajectory and its achieved objective value of (P3), denoted as $\boldsymbol{P}^{*}$ and $\tau_2^{*}$, respectively;
		\State Update $ \epsilon = |\tau_2^{(l)} - \tau_2^{*}|$, and $l = l + 1$;
		\State Update the local trajectory $\boldsymbol{P}^{(l)} = \boldsymbol{P}^{*}$ and objective value $\tau_2^{(l)}=\tau_2^{*}$;
		\EndWhile
		\EndIf
	\end{algorithmic}  
	\textbf{Output}: Optimized trajectory $\boldsymbol{P}^{(l)}$.
\end{algorithm}

\section{Numerical Results}

\begin{figure}[!t]
	\centering
	\includegraphics[scale=0.75]{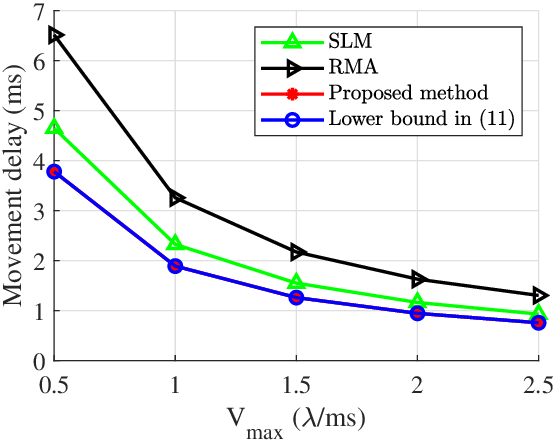}
	\setlength{\abovecaptionskip}{-0.2cm}
	\setlength{\belowcaptionskip}{-0.2cm} %调整图片标题与下文距离
	\caption{Movement delay versus maximum speed of MAs.}
	\vspace{-0.1cm}
	\label{fig:time}
\end{figure}

\begin{figure*}[htbp]
	\centering
	\vspace{-0.0cm}
	\subfigure[Initial and destination positions.]{
		\includegraphics[scale=0.43]{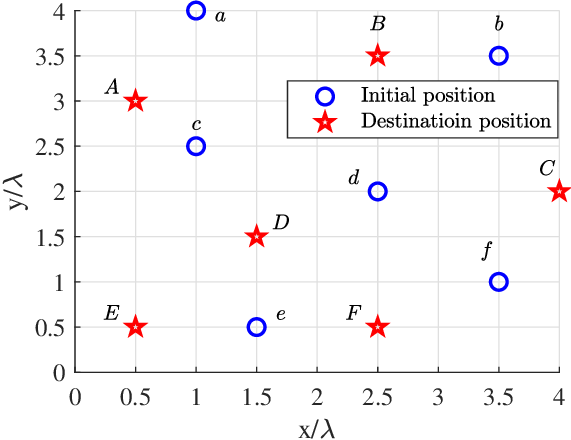}
		\label{fig:case1_path}
	}
	\subfigure[RMA.]{
		\includegraphics[scale=0.43]{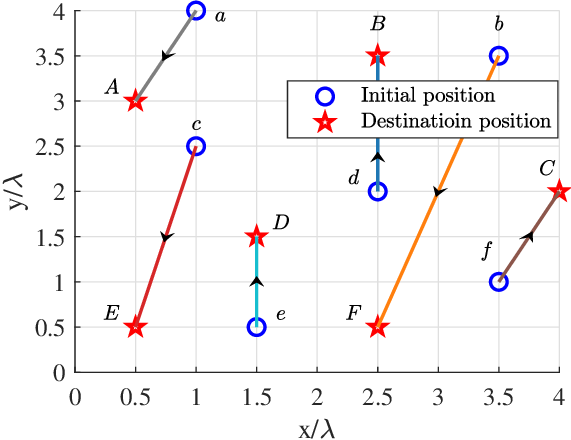}
		\label{fig:case1_path1}
	}
	\subfigure[SLM.]{
		\includegraphics[scale=0.43]{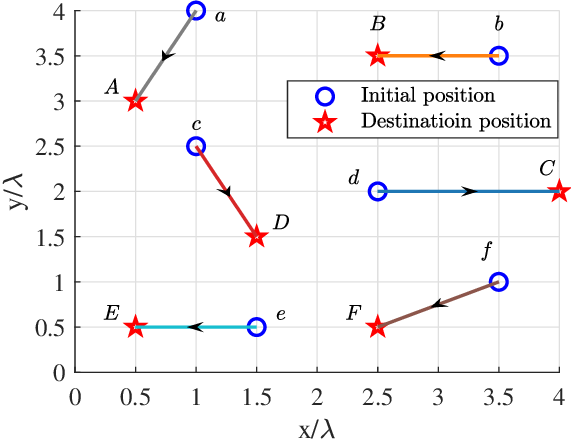}
		\label{fig:case1_path2}
	}
	\subfigure[Proposed method.]{
		\includegraphics[scale=0.43]{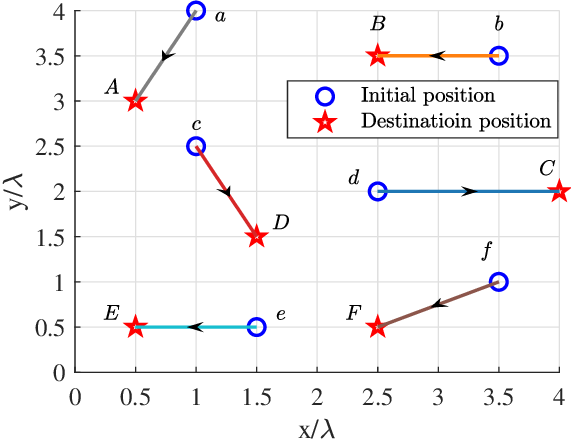}
		\label{fig:case1_path3}
	}
	
	\caption {Different trajectories for MAs in Case 1.}
	\label{fig:case1}
	\vspace{-0.0cm}
\end{figure*}

\begin{figure*}[htbp]
	\centering
	\vspace{-0.0cm}
	\subfigure[Initial and destination positions.]{
		\includegraphics[scale=0.43]{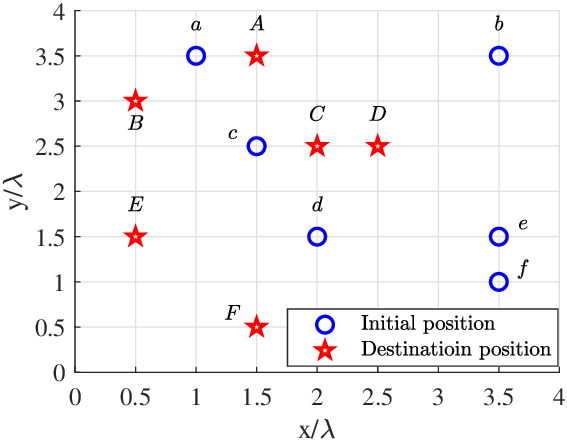}
		\label{fig:case2_path}
	}
	\subfigure[RMA.]{
		\includegraphics[scale=0.43]{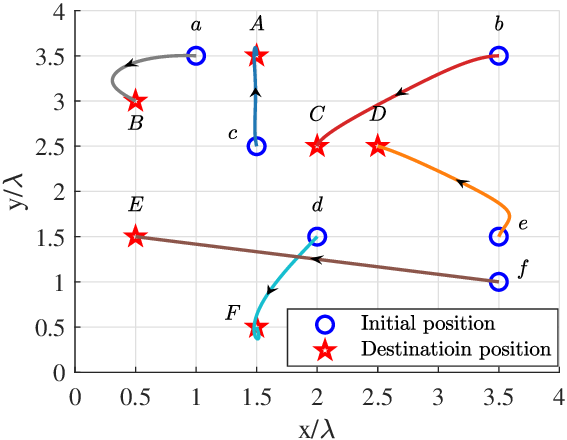}
		\label{fig:case2_path1}
	}
	\subfigure[SLM.]{
		\includegraphics[scale=0.43]{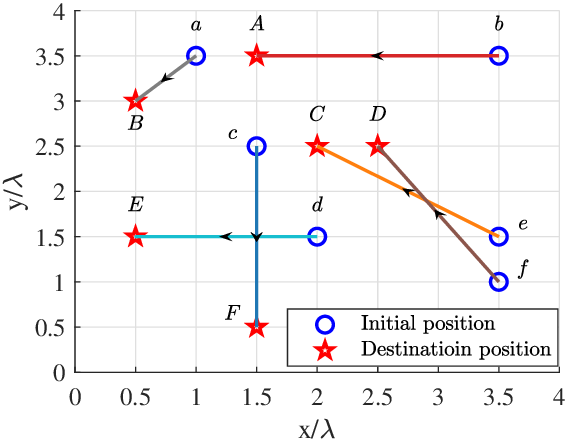}
		\label{fig:case2_path2}
	}
	\subfigure[Proposed method.]{
		\includegraphics[scale=0.43]{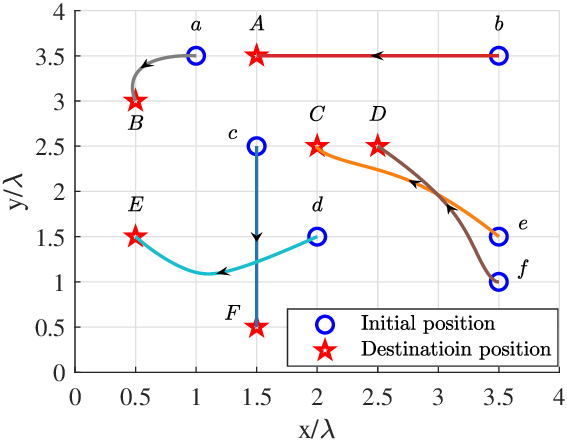}
		\label{fig:case2_path3}
	}
	
	\caption {Different trajectories for MAs in Case 2.}
	\label{fig:case2}
	\vspace{-0.0cm}
\end{figure*}
In this section, we provide numerical results to evaluate the performance of our proposed two-stage optimization algorithm for MA movement delay minimization. Unless otherwise stated, the simulation parameters are set as follows. The number of MAs is set to $M=6$ and the minimum inter-MA distance is set to $d_{\min}=\lambda/2$, where $\lambda$ is the wavelength. The transmit region is assumed to be a square region with the size of $4 \lambda \times 4 \lambda$. The total number of time slots is $N=100$. The termination threshold in Algorithm \ref{alg:SCA} is set as $\epsilon^{*}=0.001$. Furthermore, we compare the proposed algorithm with the following two benchmark schemes:
\begin{itemize}
	\item \textbf{Straight-line movement (SLM)}: Given the association solution $\sigma^*_{m,j}$'s, all MAs are constrained to move along straight lines to their destinations.

	\item \textbf{Random MA associations (RMA)}: The initial -destination association solutions are randomly generated, and the MA trajectory is optimized via SCA.
\end{itemize}

In Fig. \ref{fig:time}, we plot the movement delay versus the maximum speed of the MAs (normalized by the wavelength). The results are averaged over 100 random sets of the initial  and destination positions of MAs. It is observed from Fig. \ref{fig:time} that the movement delay decreases with the maximum speed for all of the schemes considered, as expected. In addition, the proposed two-stage algorithm yields smaller delay than the two benchmark schemes and achieves almost the same performance as the delay lower bound given in \eqref{T_lb}. It is also interesting to note that the SLM can yield lower delay compared to the random associations even with optimized trajectories. This implies that the MA position associations play a more significant role than trajectory optimization for minimizing the movement delay. Nonetheless, the SLM may not yield the optimal performance, as it requires frequent stopping and waiting in MA's movement to satisfy the inter-MA distance constrains, as will be shown later.

To gain more useful insights into the optimal MA trajectories, we further consider two specific cases as shown in Figs. \ref{fig:case1_path} and \ref{fig:case2_path} and compare the trajectories of MAs obtained by different schemes. The maximum speed of MAs is set to one wavelength per millisecond (ms).
It is observed from Figs. \ref{fig:case1_path1}, \ref{fig:case1_path2} and \ref{fig:case1_path3} that the optimized MA trajectories by all schemes in Case 1 follow straight lines, while those by the random associations differ from those by the other two schemes. This, as a result, leads to larger movement delay with random associations. Notably, the trajecotry of MA $b$ in Fig. \ref{fig:case1_path1} is observed to be much longer than those in Figs. \ref{fig:case1_path2} and \ref{fig:case1_path3}, mainly due to the lack of association optimization. 

In contrast, in Case 2, it is observed from Figs. \ref{fig:case2_path1} and \ref{fig:case2_path3} that the optimized trajectories may contain curved lines instead of straight lines only. On one hand, this is mainly due to satisfying the inter-MA distance constraints, as will be more clearly shown in Fig. \ref{fig:speed}, where we plot the speeds of MAs $b$, $e$, and $f$ over time by the SLM benchmark and the proposed method. It is observed that with curved paths, the speeds of these MAs can remain close to the maximum speed. However, under the SLM, the speeds of MAs $e$ and $f$ need to be much lower than the maximum speed over certain periods. In particular, MA $e$ initially moves at the maximum speed but gradually decelerates, while MA $f$ initially moves at a low speed to avoid collision with MA $e$. This move-wait-move trajectory results in even longer delay compared to detours in the trajectories obtained by our proposed algorithm.

On the other hand, the curved paths also imply the non-unique optimal trajectory solutions to (P1). For example, for the MA initial-destination position pair $(a,B)$, even MA $a$ follows a curved trajectory to $B$, its movement time is still not the longest among all MAs. Thus, MA $a$ can have many options for its trajectory as long as its movement time is not the longest among all MAs. 

\begin{figure}[!t]
	\centering
	\includegraphics[scale=0.75]{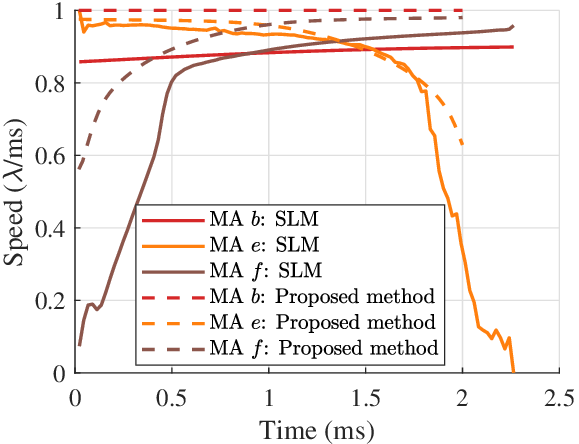}
	\setlength{\abovecaptionskip}{-0.2cm}
	\setlength{\belowcaptionskip}{-0.2cm} %调整图片标题与下文距离
	\caption{Speeds over time for MAs by different schemes in Case 2.}
	\vspace{-0.1cm}
	\label{fig:speed}
\end{figure}

\section{Conclusion}

In this paper, we studied a new trajectory optimization problem for MAs to minimize their movement delay subject to practical inter-MA distance constraints. A two-stage optimization algorithm is proposed to first solve a simplified problem without the inter-MA distance constraints to obtain the MA position associations and then construct a feasible trajectory solution via SCA if needed. Numerical results demonstrate that our proposed algorithm can achieve near-optimal performance in terms of movement delay minimization, and the MA position associations play a significant role in reducing the overall delay. Furthermore, moving all MAs along straight lines may result in large delay due to the long waiting time for MA collision avoidance.

\bibliographystyle{IEEEtran}
\bibliography{IEEEabrv,reference_abrv}

\end{document}